\documentclass[letterpaper, 10 pt, conference]{settings/ieeeconf}
\IEEEoverridecommandlockouts
\title{\LARGE \bf 
    Formal Verification of Linear Temporal Logic Specifications \\ Using Hybrid Zonotope-Based Reachability Analysis
    }

\author{Loizos Hadjiloizou$^{1}$, Frank J. Jiang$^{1}$, Amr Alanwar$^{2}$, Karl H. Johansson$^{1}$
\thanks{This work was partially supported by the Wallenberg Artificial Intelligence, Autonomous Systems, and Software Program (WASP) funded by the Knut and Alice Wallenberg Foundation. It was also partially supported by the Swedish Research Council, Swedish Research Council Distinguished Professor Grant 2017-01078, and the Knut and Alice Wallenberg Foundation Wallenberg Scholar Grant.}
\thanks{$^{1}$L. Hadjiloizou, F. J. Jiang, and K. H. Johansson are with the Division of Decision and Control Systems, EECS, KTH Royal Institute of Technology, Malvinas väg 10, 10044 Stockholm, Sweden,
        {\tt\small \{loizosh, frankji, kallej\}@kth.se}. They are also affiliated with Digital Futures.}
\thanks{$^{2}$A. Alanwar is with the School of Computation, Information and Technology, Technical University of Munich, Bildungscampus 2, 74076 Heilbronn, Germany,  {\tt\small alanwar@tum.de}}}

\usepackage{amsmath}
\usepackage{mathdots}
\usepackage{yhmath}
\usepackage{siunitx}
\usepackage{array}
\usepackage{amssymb}
\usepackage{mathtools}                  
\newcommand{\vect}[1]{\boldsymbol{#1}}  

\usepackage{graphicx}
\usepackage{subcaption}

\usepackage{algorithm}
\usepackage{algpseudocode}

\usepackage{pifont}
\newcommand{\cmark}{\ding{51}}%
\newcommand{\xmark}{\ding{55}}%

\usepackage{makecell}

\usepackage{balance}
\newtheorem{definition}{Definition}[section]
\newtheorem{example}{Example}[section]
\newtheorem{problem}{Problem}[section]

\usepackage{comment}

\usepackage{soul}

\begin{document}
    \maketitle
    \thispagestyle{empty}
    \pagestyle{empty}

    \begin{abstract}

In this paper, we introduce a hybrid zonotope-based approach for formally verifying the behavior of autonomous systems operating under Linear Temporal Logic (LTL) specifications. In particular, we formally verify the LTL formula by constructing temporal logic trees (TLT)s via backward reachability analysis (BRA). In previous works, TLTs are predominantly constructed with either highly general and computationally intensive level set-based BRA or simplistic and computationally efficient polytope-based BRA. In this work, we instead propose the construction of TLTs using hybrid zonotope-based BRA. By using hybrid zonotopes, we show that we are able to formally verify LTL specifications  in a computationally efficient manner while still being able to represent complex geometries that are often present when deploying autonomous systems, such as non-convex, disjoint sets. Moreover, we evaluate our approach on a parking example, providing preliminary indications of how hybrid zonotopes facilitate computationally efficient formal verification of LTL specifications in environments that naturally lead to non-convex, disjoint geometries.

\end{abstract}
    \section{INTRODUCTION}
\label{sec:introduction}

Advancements in fields like computer science, hybrid systems, and cyber-physical systems have ushered in a new era of complexity. As we extend the limits of what's possible, ensuring the safety of ever more complex systems becomes crucial. This endeavor requires adept methods for describing and analyzing their complexity. A promising approach is to leverage temporal logic formalisms~\cite{Baier2008Principles} to capture system objectives and then use reachability analysis~\cite{Borelli2017Predictive} to study its behaviour.

A significant body of literature employing this approach, focuses on automata-based methods~\cite{Baier2008Principles},~\cite{Kloetzer2008Fully}. While powerful, automata tend to suffer from computational and expressive limitations when applied to continuous state-space systems with time-varying specifications or environments~\cite{Gao2021Temporal}. In this work, we further explore an alternative to automata first proposed in~\cite{Gao2021Temporal}, based on a tree structure called Temporal Logic Tree to encode information about the satisfaction of an LTL specification. Unlike automata, they are abstraction-free for continuous state-space systems, support the full LTL language, and are more modular, enabling online adaption to time-varying specifications or environments.

The computational complexity of constructing a TLT depends largely on the chosen BRA technique. In~\cite{Jiang2020Ensuring}, Hamilton-Jacobi reachability analysis is used to build a TLT, ensuring safety in vehicle parking tasks. This method facilitates the construction of a TLT for systems with nonlinear dynamics, and through a level-set representation, for environments with intricate geometries. While powerful and general, Hamilton-Jacobi reachability analysis' computational complexity grows exponentially with the system's state space~\cite{Chen2018}. More computationally efficient tools for reachability analysis do exist, such as polytope, zonotope $(\mathcal{Z})$, constrained zonotope $(\mathcal{CZ})$, sparse polynomial zonotope $(\mathcal{SPZ})$, and constrained polynomial zonotope $(\mathcal{CPZ)}$-based methods~\cite{Herceg2013, Girard2005Reachability, Scott2016Constrained, Kochdumper2020, Kochdumper2023Constrained}. While these approaches can represent geometries of varying complexity, to the extent of our knowledge, they can not efficiently handle disjoint sets, which is often crucial for autonomous systems. A recent work proposed a new set representation called hybrid zonotope $(\mathcal{HZ})$ that, among other benefits, can efficiently handle disjoint sets~\cite{Bird2023Hybrid}. By leveraging $\mathcal{HZ}$-based BRA in this work, we aim to efficiently verify LTL specifications that include disjunction operators in, for example, a parking lot environment where the reachable sets naturally become disjoint~\cite{Jiang2020Ensuring}.
\begin{figure}[t]
    \centering
    \includegraphics[width=0.90\columnwidth, height=0.42\columnwidth]{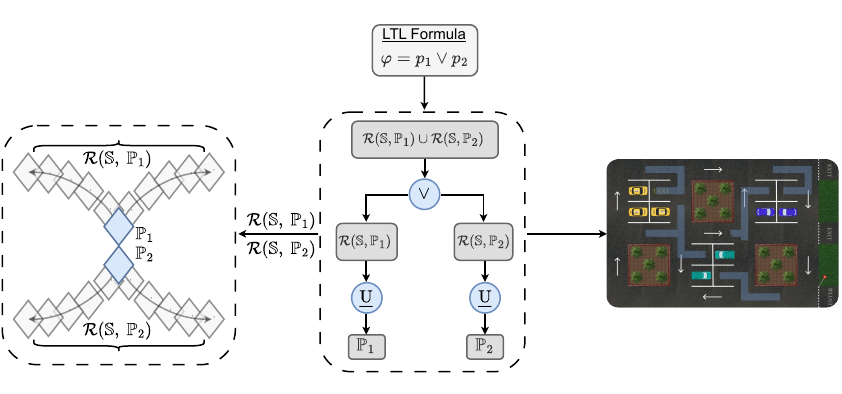}
    \caption{We illustrate an overview of our approach.} 
    \label{fig:block_diagram}
    \vspace{-0.7cm}
\end{figure}

The main contribution of this work is a verification approach that uses $\mathcal{HZ}$-based BRA to construct TLTs in a computationally efficient-manner. This allows for the efficient verification of the feasibility of any LTL formula in environments with relatively challenging geometries. More specifically, the contributions are outlined as follows:
(1) detailing the construction of TLTs using $\mathcal{HZ}$-based BRA, (2) implementing the proposed method, which is publicly available at \footnote{https://github.com/loizoshad/zonopy}, and (3) evaluating the implemented approach on a practical example with an LTL specification containing disjunction operators and disjoint reachable sets.

The remainder of this paper is structured as follows: We introduce preliminary material on reachability analysis, LTL, and TLTs in Section~\ref{sec:preliminaries}. Section~\ref{sec:problem_statement} formally introduces the problem solved in this work as well as provides a motivation for choosing $\mathcal{HZ}$s.  Section~\ref{sec:constructing_tlts} lays out our approach for constructing a TLT using $\mathcal{HZ}$s. In Section~\ref{sec:simulated_parking_scenario}, we formulate a parking case study. Section~\ref{sec:results} analyzes the results from our simulations and finally, Section~\ref{sec:conclusion}, concludes our work and discusses some future work that can be done in this area.
    \section{PRELIMINARIES}
\label{sec:preliminaries}

\subsection{Notation}

We denote the set of real numbers as $\mathbb{R}$, vectors as bold lowercase letters (e.g., $\vect{x} \in \mathbb{R}^{n}$), the identity matrix by $I$, while matrices filled with elements $1$ and $0$ are represented by $\vect{1}, \vect{0}$, respectively. We extract a submatrix from $G$ by selecting its first $n$ rows as $G[1:n,:]$ and express the horizontal concatenation of matrices $A$ and $B$ as $[A~B]$. A diagonal matrix is denoted by $\textit{diag}(\vect{v})$, with $\vect{v}$ being the vector with the diagonal elements. The $n$-dimensional and constrained n-dimensional unit hypercubes are denoted by $\mathcal{B}_{\infty}^{n} = \{ \vect{\xi} \in \mathbb{R}^{n} | ~||\vect{\xi}||_{\infty} \leq 1 \}$ and $\mathcal{B}_{\infty}^{n}(A, \vect{b}) =\{ \vect{\xi} \in \mathbb{R}^{n} | ~||\vect{\xi}||_{\infty} \leq 1, A\vect{\xi} = \vect{b} \}$, respectively, while the power set of an $n$-dimensional vector of binary variables as $\{ -1, 1 \}^{n}$.

\subsection{Plant Model}

Consider the discrete-time system
\begin{align}
    \vect{x}_{t+1} = A\vect{x}_{t} + B\vect{u}_{t},
    \label{chapter_preliminaries:eq:dynamics_model}
\end{align}
where $\vect{x}_{t} \in \mathbb{R}^{n_{x}}$ represents the state vector, $\vect{u}_{t} \in \mathbb{R}^{n_{u}}$ the control input, and the matrices $A\in\mathbb{R}^{n_{x} \times n_{x}}, B\in\mathbb{R}^{n_{x} \times n_{u}}$, are the state transition and control input matrices, respectively. For each time instant $t$, $u_{t} \in \mathbb{U} \subset \mathbb{R}^{n_{u}}$, where $\mathbb{U}$ is the compact control input constraint. We assume the system is well-defined such that $x_{t+1}$ is uniquely determined by $x_t$ and $u_t$. Let $\mu = u_0, u_1, \ldots$ be a control policy and $\mathcal M$ be the set of all control policies. Then, given initial state $x_0$, we denote a trajectory of system~\eqref{chapter_preliminaries:eq:dynamics_model} as $\zeta(\cdot; x_0, \mu)$ where $\zeta(t; x_0, \mu)$ is the state of system~\eqref{chapter_preliminaries:eq:dynamics_model} at time $t$ when starting from $x_0$ and implementing control policy $\mu$. For simplicity, we will sometimes refer to trajectories with $\zeta(\cdot)$.

\subsection{Reachability Analysis}

Reachability analysis is a common approach to formally provide guarantees about the behavior of a system. In this subsection, we provide the relevant reachability definition that will later be used for the construction of TLTs.

\begin{definition}
    Consider the plant~\eqref{chapter_preliminaries:eq:dynamics_model}, then the infinite horizon backward reachable set (BRS) from the target set $\mathcal{T} \subseteq \mathbb{R}^{n_x}$ in state space $\mathbb{S} \subseteq \mathbb{R}^{n_x}$ is given as
    \begin{equation}
        \begin{split}
            \mathcal{R}(\mathbb{S}, \mathcal{T}) = 
            \{
            & x \in \mathbb{S}| \exists \mu \in \mathcal{M}, \exists t > 0, \zeta(t; x, \mu) \in \mathcal{T}
            \}.
        \end{split}
    \end{equation}
\end{definition}
Intuitively, the infinite horizon BRS answers the question of where the system can begin and eventually reach the target set $\mathcal{T}$. This definition is in accordance with the controlled reachable set definition in~\cite{Gao2021Temporal}. As it will become apparent later, it allows us to construct the controlled TLT, which for the sake of brevity, we generally refer to as TLT in this work.

In practice, to compute the infinite horizon BRS, one needs to compute the union of all N-step BRSs for $N = [0, \infty), N \in \mathbb{Z}$, where the N-step BRS from the target set $\mathcal{T} \subseteq \mathbb{R}^{n_x}$ in $\mathbb{S} \subseteq \mathbb{R}^{n_x}$ is the set of all states for which system~\eqref{chapter_preliminaries:eq:dynamics_model} can reach the target set in exactly N time steps. The N-step BRS is equivalent to computing the predecessor set N times,
(e.g., $\mathcal{R}(\mathbb{S}, \mathcal{T}, 3) = \mathcal{P}(\mathbb{S}, \mathcal{P}(\mathbb{S}, \mathcal{P}(\mathbb{S}, \mathcal{T})))$)
which in turn is defined as
\begin{equation}
    \begin{split}
        \mathcal{P}(\mathbb{S}, \mathcal{T}) = 
        \{
        & x \in \mathbb{S}| \exists u \in \mathbb{U}, Ax + Bu \in \mathcal{T}
        \}.        
    \end{split}
    \label{eq:predecessor_set}
\end{equation}

\subsection{Linear Temporal Logic}

In the context of this work, LTL plays a pivotal role in modeling complex temporal tasks to study the temporal and time-invariance properties of autonomous systems. At its core, an LTL formula consists of three components: a finite set of atomic propositions ($\mathsf{P}$), a set of temporal ($until: \underline{\textit{U}}$), and logical ($negation \coloneqq \lnot,~and \coloneqq \land$) operators. In this work, we consider the following LTL syntax:
\begin{align}
    \varphi ::= ~ true ~ | ~ p \in \mathsf{P} ~ | ~ \neg \phi\ ~ | ~ \phi_{1} \land \phi_{2}\ ~ | ~ \phi_{1} \underline{\textit{U}} \phi_{2} ~ | ~ \bigcirc \phi.
    \label{eq:preliminaries:ltl_syntax}
\end{align}
\begin{definition}
    For an LTL formula $\varphi$, a labeling function $l(\cdot)$, a trajectory $\zeta(\cdot)$, and a time instant $t \geq 0$, the satisfaction relation $(\zeta(\cdot), t) \vDash \varphi$ is defined as
    \begin{equation}
        \begin{split}
            (\zeta(\cdot), t) & \vDash p \in \mathsf{P} \Leftrightarrow \zeta(t) \in l(p), \\
            (\zeta(\cdot), t) & \vDash \neg \varphi \Leftrightarrow  (\zeta(\cdot), t) \nvDash \varphi, \\
            (\zeta(\cdot), t) & \vDash \varphi_{1} \land \varphi_{2} \Leftrightarrow (\zeta(\cdot), t) \vDash \varphi_{1} \land (\zeta(\cdot), t) \vDash \varphi_{2},   \\
            (\zeta(\cdot), t) & \vDash \varphi_{1} \lor \varphi_{2} \Leftrightarrow (\zeta(\cdot), t) \vDash \varphi_{1} \lor (\zeta(\cdot), t) \vDash \varphi_{2},   \\
            (\zeta(\cdot), t) & \vDash \lozenge \varphi \Leftrightarrow \exists t_{1} \in [t,\infty) : (\zeta(\cdot), t_{1}) \vDash \varphi, \\
            (\zeta(\cdot), t) & \vDash \square \varphi \Leftrightarrow \forall t_{1} \in [t,\infty) : (\zeta(\cdot), t_{1}) \vDash \varphi, \\
            (\zeta(\cdot), t) & \vDash \varphi_{1} \underline{\textit{U}} \varphi_{2} \Leftrightarrow \exists \ t_{1} \in [t, \infty) :  (\zeta(\cdot), t_{1}) \vDash \varphi_{2}, \\
            & \hspace{0.44cm} \forall t_{2} \in [t, t_{1}), (\zeta(\cdot), t_{2}) \vDash \varphi_{1}, \\
            (\zeta(\cdot), t) & \vDash \bigcirc \varphi \Leftrightarrow (\zeta(\cdot), t+1) \vDash \varphi
            .
        \end{split}
    \end{equation}
\end{definition}

\smallskip

\subsection{Temporal Logic Tree}
A TLT is a hierarchical structure that describes the satisfaction relationship between the system's state space and LTL specification. This structure serves as an alternative approach for model checking and control synthesis, which provides high modularity. Another significant characteristic of the TLT is that it enables the synthesis of a controller with multiple control policies, which is useful for applications with mixed levels of autonomy, such as autonomous vehicles.
\begin{definition}
    A TLT is a tree for which
    \begin{itemize}
        \item each node is either a set node within $\mathbb{R}^{n_{x}}$ or an operator node from the LTL syntax defined in~(\ref{eq:preliminaries:ltl_syntax});
        \item the root and leaf nodes are set nodes;
        \item a set node that is not a leaf node has as a unique child an operator node;
        \item all operator nodes have set nodes are their children.
    \end{itemize}
\end{definition}
\begin{figure}[b]
	\begin{center}
            \vspace{-0.3cm}
		\includegraphics[width=\columnwidth, height=0.55\columnwidth]{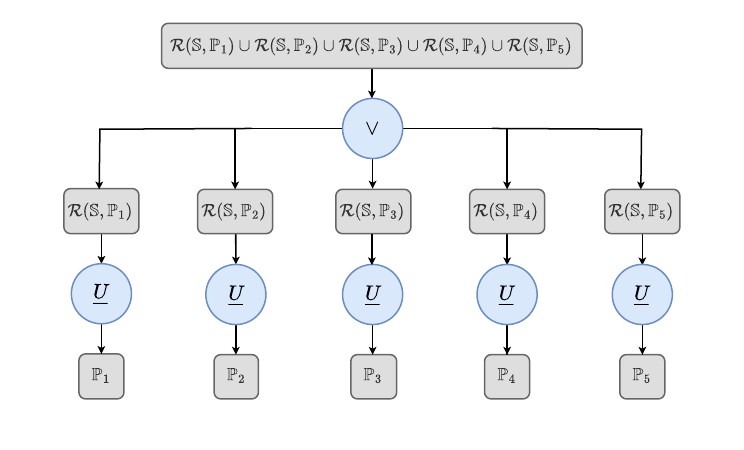}
		\caption{TLT to check if the vehicle can eventually park.}
		\label{fig:example_tlt}
            \vspace{-0.5cm}
	\end{center}
\end{figure}
\begin{example}\label{example:tlt}
Given the dynamic model of an autonomous vehicle described by~\eqref{chapter_preliminaries:eq:dynamics_model}, study the eligibility of all states to allow the vehicle to eventually reach one of the available parking spots ($\mathbb{P}_{1}, \mathbb{P}_{2}, ..., \mathbb{P}_{5}$). Given this description, we can formulate the following LTL to capture this goal.
\begin{equation}
    \begin{split}
        \varphi = (\lozenge p_{1}) \lor (\lozenge p_{2}) \lor (\lozenge p_{3}) \lor (\lozenge p_{4}) \lor (\lozenge p_{5}), \\
        p_{i} = \{ \vect{x} \in \mathbb{S} | \vect{x} \in \mathbb{P}_{i} \}, \hspace{0.3cm} 
        i = \{1, 2, ..., 5\}. \hspace{0.35cm}
    \end{split}
    \label{eq:ltl_example}
\end{equation}
\begin{itemize}
    \item \textbf{Step 1}: Obtain weak-until positive normal form (PNF) \cite{Baier2008Principles}. The PNF for LTL permits negation only on the level of atomic propositions. To ensure the full expressiveness of LTL, in addition to the LTL fragment in~\eqref{eq:preliminaries:ltl_syntax} it includes the Boolean \textit{Or} $\lor$ operator as well us the \textit{Weak-Until} $\underline{\textit{W}}$ operator, where $\varphi_{1} \underline{\textit{W}} \varphi_{2} \equiv \varphi_{1} \underline{\textit{U}} \varphi_{2} \lor \square \varphi_{1} $.    
    \begin{align}
        \varphi_{N} = (true \, \underline{\textit{U}} \, p_{1}) \lor \dots \lor (true \, \underline{\textit{U}} \,  p_{5}).
        \label{eq:ltl_example_pnf}
    \end{align}
    \item \textbf{Step 2}: Compute the TLT for each atomic proposition. The TLT for any atomic proposition is simply a set node consisting of all states that satisfy it. In this example, these nodes are the set nodes, each containing the equivalent parking area $\mathbb{P}_{i}$ for every parking spot $i$. These are the leaf nodes of the TLT in Figure~\ref{fig:example_tlt}.    
    \item \textbf{Step 3}: Inductively construct the TLT for each component of the formula. Each branch in Figure~\ref{fig:example_tlt} represents the TLT for each component $(true \, \underline{\textit{U}} \, p_{i})$. Every branch consists of two set nodes and one operator node. The node at the bottom is the target $i$ the vehicle needs to reach, while the top set node is the set of all states that could eventually take the vehicle to target $i$, hence the backward reachable set of the target $\mathbb{P}_{i}$. These two nodes are connected by the \textit{Until} temporal operator. Intuitively, each branch asks that the vehicle is within the BRS of each target until it reaches the target itself. After combining all the branches' backward reachable sets through set unions, the root node consists of all the states for which we can guarantee satisfaction of the LTL specification.    
\end{itemize}
\end{example}

This example hints that the effectiveness of TLTs in formally verifying LTL specifications relies on the set representation's support for different operators.

    \section{Problem Statement}
\label{sec:problem_statement}

Constructing a TLT for any LTL problem involves computations such as BRSs, intersections, and unions. In many cases, this results in non-convex, disjoint sets. While level-set Hamilton Jacobi reachability can handle such complexities, it comes with increased computational expense. Conversely, more computationally efficient methods can construct TLTs in relatively simple scenarios but to the best of our knowledge, can not efficiently handle the full LTL language requirements. Table~\ref{table:set_representation_tools} provides an overview of common set representations for reachability analysis and their suitability for addressing these requirements, among which, only level sets and $\mathcal{HZ}$s can effectively handle the full LTL language. Due to the computational complexity of level-set solutions, we explore the use of $\mathcal{HZ}$s in this work and define the following problem statement.

\begin{table}[t]
    \vspace{0.4cm}
    \centering
    \begin{tabular}{|@{}c@{}|@{}c@{}|@{}c@{}|@{}c@{}|@{}c@{}|@{}c@{}|@{}c@{}|@{}c@{}|}    
        \hline
        \textbf{} & ~Polytope~ & ~$\mathcal{Z}$~ & ~$\mathcal{CZ}$~ & ~$\mathcal{SPZ}$~ & ~$\mathcal{CPZ}$~ & ~$\mathcal{HZ}$~ & \makecell{\hspace{0.1cm}Level\:\\Set} \\
        \hline
        \makecell{Linear\\~Transformation\,} & \cmark & \cmark & \cmark & \cmark & \cmark & \cmark & \cmark \\
        \hline
        \makecell{Minkowski\\Sum} & \cmark & \cmark & \cmark & \cmark & \cmark & \cmark & \cmark \\
        \hline
        Intersection & \cmark & \xmark & \cmark & \xmark & \cmark & \cmark & \cmark \\
        \hline
        Union & \xmark & \xmark & \xmark & \xmark & \cmark & \cmark & \cmark \\
        \hline
        Non-Convex & NO & NO & NO & YES & YES & YES & YES \\
        \hline
        Disjoint & NO & NO & NO & NO & NO & YES & YES \\
        \hline
    \end{tabular}
    \caption{Properties of set representation tools\protect\footnotemark}
    \label{table:set_representation_tools}
    \vspace{-0.2cm}
\end{table}
\footnotetext{\cmark: Set representation closed under the operation, \xmark: Set representation not closed under the operation}

\begin{problem}
    For an autonomous system with the dynamics~\eqref{chapter_preliminaries:eq:dynamics_model} and an LTL specification $\varphi$, formally study the satisfiability of $\varphi$ by~\eqref{chapter_preliminaries:eq:dynamics_model} in a computationally efficient manner for environments that naturally form non-convex, disjoint sets. 
    \label{problem}
\end{problem}
    \section{Constructing TLTs using Hybrid Zonotopes}
\label{sec:constructing_tlts}

In this section, we introduce and formulate our approach for constructing TLTs using hybrid zonotopes.

\subsection{Hybrid Zonotopes}

To understand the $\mathcal{HZ}$, let's start by defining its predecessor, the zonotope which is a centrally symmetric convex polytope, defined as the affine image of a unit hypercube.

\begin{definition} 
    A set $\mathcal{Z} \subset \mathbb{R}^{n_{x}}$ is a zonotope if there exist a matrix $G \in \mathbb{R}^{n_{x} \times n_{g}}$ and a vector $\vect{c} \in \mathbb{R}^{n_{x}}$, where $n_{g}$ the number of generators, such that $\mathcal{Z} = \{ \vect{c} + G\vect{\xi} ~|~ ||\vect{\xi}||_{\infty} \leq 1\}$.
\end{definition}

The constrained zonotope is an immediate extension of the zonotope, which allows linear equality constraints to be imposed on its coefficients $\vect{\xi}$. This results in a non-centrally symmetric convex polytope.

\begin{definition}
    A set $\mathcal{CZ} \subset \mathbb{R}^{n_{x}}$ is a constrained zonotope if there exist matrices $G \in \mathbb{R}^{n_{x} \times n_{g}}, A \in \mathbb{R}^{n_{c} \times n_{g}}$ and vectors $\vect{c} \in \mathbb{R}^{n_{x}}, \vect{b} \in \mathbb{R}^{n_{c}}$, where $n_{c}$ is the number of linear equality constraints, such that $\mathcal{CZ} = \{ \vect{c} + G\vect{\xi} ~|~ ||\xi||_{\infty} \leq 1, A\vect{\xi} = \vect{b} \}$.
\end{definition}    

The constrained zonotope is represented in its constrained generator form and, in short, denoted as $\mathcal{CZ} = \{\vect{c}, G, A, \vect{b}\}$.

The hybrid zonotope extends the notion of the constrained zonotope by restricting the coefficients of some generators to lie only on the edges of the unit hypercube. These generators are called the binary generators of the $\mathcal{HZ}$. This can create gaps in the set due to the absence of entries from the remaining coefficients in the range $(-1, 1)$. Intuitively, a $\mathcal{HZ}$ is equivalent to the union of $2^{n_{b}}$ constrained zonotopes~\cite{Bird2023Hybrid}, where $n_{b}$ is the number of binary generators of the $\mathcal{HZ}$.

\begin{definition}
    A set $\mathcal{HZ} \subset \mathbb{R}^{n_{x}}$ is a hybrid zonotope if there exist matrices $G^{c} \in \mathbb{R}^{n_{x} \times n_{g}}, G^{b} \in \mathbb{R}^{n_{x} \times n_{b}}, A^{c} \in \mathbb{R}^{n_{c} \times n_{g}}, A^{b} \in \mathbb{R}^{n_{c} \times n_{b}}$ and vectors $\vect{c} \in \mathbb{R}^{n_{x}}, \vect{b} \in \mathbb{R}^{n_{c}}$ such that
    \begin{equation}
        \begin{split}
            \mathcal{HZ} = 
            \bigg\{ 
            \begin{bmatrix} G^{c} & G^{b}  \end{bmatrix} 
            \begin{bmatrix} \vect{\xi}^{c} \\ \vect{\xi}^{b} \end{bmatrix} + 
            \vect{c} \bigg|
            & \begin{bmatrix} \vect{\xi}^{c} \\ \vect{\xi}^{b} \end{bmatrix} \in 
            \mathcal{B}_{\infty}^{n_{g}} \times \{-1, 1\}^{n_{b}}, \\
            & \begin{bmatrix} A^{c} & A^{b} \end{bmatrix} 
            \begin{bmatrix} \vect{\xi}^{c} \\ \vect{\xi}^{b} \end{bmatrix} = 
            \vect{b} 
            \bigg\}.
        \end{split}
    \end{equation}
\end{definition}
The hybrid zonotope is represented in its hybrid constrained generator form and is, in short, denoted as $\mathcal{HZ} = \{\vect{c}, G^{c}, G^{b}, A^{c}, A^{b}, \vect{b}\}$.

\subsection{TLT Construction}

Now, it is time to show how one can use the $\mathcal{HZ}$ to verify whether an LTL specification can be satisfied. We start by defining an LTL formula whose feasibility we want to verify. Then we obtain its PNF and define an $\mathcal{HZ}$ for each of its atomic propositions such that it consists all the states that satisfy that particular atomic proposition. Then, we inductively construct each sub-TLT corresponding to a temporal or logical operation. Starting with the logical \textit{Or} and \textit{And} operations, the construction of their TLTs requires the computation of the union and the intersection of two $\mathcal{HZ}s$ respectively. Meanwhile, the construction of a TLT for the \textit{Until} and \textit{Weak-Until} operators requires the computation of unions, backward reachable sets, as well as robust controlled invariant (RCI) sets for the latter operation. As shown in~\cite{Borelli2017Predictive}, the computation of an RCI is a series of predecessor and intersection set operations. Following Table~\ref{table:constructing_tlts}, the TLT for any LTL formula can be constructed. In Table~\ref{table:constructing_tlts}, $\mathcal{HZ}_{\scriptscriptstyle \phi}$ is the computed set for satisfying sub-formula $\phi$ and $\mathcal{RCI}(\phi) = \cap_{\scriptscriptstyle i = 0}^{\scriptscriptstyle \infty} \mathcal{R}(\mathbb{S}, \mathcal{HZ}_{\scriptscriptstyle \phi})$.

\begin{table}[t]
\vspace{0.2cm}
\begin{center}
    \caption{Corresponding operation for the LTL syntax}
    \begin{tabular}{|c|c|}
        \hline
        \textbf{LTL} & \textbf{Set Operation for TLT Construction} \\
        \hline 
        $p \in \mathsf{P}$  & $\mathcal{HZ}_{\scriptscriptstyle p \in \mathsf{P}} = \{ s \in \mathbb{P} \}$ \\
        \hline 
        $true$ & $\mathcal{HZ} = \{ s \in \mathbb{S} \}$ \\
        \hline 
        $\neg \phi$ & $\mathcal{HZ} = \textit{compl}(\mathcal{CZ}),$~\cite[eq. (16)]{Bird2021Unions} \\        
        \hline 
        $\phi_{\scriptscriptstyle 1} \land \phi_{\scriptscriptstyle 2}$ & $\mathcal{HZ}_{\scriptscriptstyle \phi_{\scriptscriptstyle 1} \land \phi_{\scriptscriptstyle 2}} = \textit{inters}(\mathcal{HZ}_{\scriptscriptstyle \phi_{\scriptscriptstyle 1}}, \mathcal{HZ}_{\scriptscriptstyle \phi_{\scriptscriptstyle 2}}),$~\cite[eq. (8)]{Bird2023Hybrid} \\
        \hline 
        $\phi_{\scriptscriptstyle 1} \lor \phi_{\scriptscriptstyle 2}$ & $\mathcal{HZ}_{\scriptscriptstyle \phi_{\scriptscriptstyle 1} \lor \phi_{\scriptscriptstyle 2}} = \textit{union}(\mathcal{HZ}_{\scriptscriptstyle \phi_{\scriptscriptstyle 1}}, \mathcal{HZ}_{\scriptscriptstyle \phi_{\scriptscriptstyle 2}}),$~\cite[eq. (4)]{Bird2021Unions} \\
        \hline 
        $\phi_{\scriptscriptstyle 1} \underline{\textit{U}} \phi_{\scriptscriptstyle 2}$ & $\mathcal{HZ}_{\scriptscriptstyle \phi_{\scriptscriptstyle 1} \underline{\textit{U}} \phi_{\scriptscriptstyle 2}} = \mathcal{R}(\mathcal{HZ}_{\scriptscriptstyle \phi_{\scriptscriptstyle 1}}, \mathcal{HZ}_{\scriptscriptstyle \phi_{\scriptscriptstyle 2}})$ \\
        \hline 
        $\phi_{\scriptscriptstyle 1} \underline{\textit{W}} \phi_{\scriptscriptstyle 2}$ & $\mathcal{HZ}_{\scriptscriptstyle \phi_{\scriptscriptstyle 1} \underline{\textit{W}} \phi_{\scriptscriptstyle 2}} = \textit{union}(\mathcal{RCI}(\mathcal{HZ}_{\scriptscriptstyle \phi_{\scriptscriptstyle 1}}), \mathcal{HZ}_{\scriptscriptstyle \phi_{\scriptscriptstyle 2}})$ \\
        \hline 
        $\lozenge \phi = \textit{true} \underline{\textit{U}}\phi$ & $\mathcal{HZ}_{\scriptscriptstyle \textit{true} \underline{\textit{U}} \phi} = \mathcal{R}(\mathbb{S}, \mathcal{HZ}_{\scriptscriptstyle \phi})$ \\
        \hline 
        $\square \phi = \phi \underline{\textit{W}} \textit{false}$ & $\mathcal{HZ}_{\scriptscriptstyle \phi \underline{\textit{W}} \textit{false}} = \mathcal{RCI}(\mathcal{HZ}_{\scriptscriptstyle \phi})$ \\
        \hline 
        $\bigcirc \phi$ & $\mathcal{HZ}_{\scriptscriptstyle \bigcirc\phi} = \mathcal{P}(\mathbb{S}, \mathcal{HZ}_{\scriptscriptstyle \phi})$ \\
        \hline
    \end{tabular}
    \label{table:constructing_tlts}
\end{center}
\end{table}

\begin{algorithm}
    \caption{Constructing TLT using $\mathcal{HZ}$}\label{alg:construct_tlt}
    \begin{algorithmic}[1]
        \State \textbf{\small Input:} \hspace{0.07cm} $\varphi_{N}$ \Comment{\footnotesize LTL specification~~\:}
        \State \hspace{1.1cm} $\hat{\mathbb{S}} = \{ \vect{c}_{s}, G_{s}^{c}, G_{s}^{b}, A_{s}^{c}, A_{s}^{b}, \vect{b}_{s} \}$, \Comment{\footnotesize Augm. State Space~~}
        \State \hspace{1.1cm} $\mathbb{P}_{1} = \{ \vect{c}_{t}, G_{t}^{c}, G_{t}^{b}, A_{t}^{c}, A_{t}^{b}, \vect{b}_{t} \}$, \Comment{\footnotesize Target Space ~~~~~~~\:}
        \State \hspace{1.1cm} $D = [A ~ B]$, \Comment{\footnotesize System Dynamics~~\:}
        \State \textbf{\small Output:} $\mathcal{R}(\hat{\mathbb{S}}, \mathbb{P}_{1})$ \Comment{\footnotesize Root Set Node~~~~~\:\:}
        \State \textbf{\small Process:}
        \State $\mathcal{R}(\hat{\mathbb{S}}, \mathbb{P}_{1}) \gets \mathbb{P}_{1}$ \Comment{\footnotesize Init BRS~~~~~~~~~~~~~~}
        \State $Pre \gets \mathbb{P}_{1}$ \Comment{\footnotesize Init Predecessor Set}
        \While{$\text{True}$}
            \State $Pre \gets \mathcal{P}(\hat{\mathbb{S}}, Pre)$ \Comment{\footnotesize Predecessor Set~~~~~~}
            \If{$Pre \subseteq \mathcal{R}(\hat{\mathbb{S}}, \mathbb{P}_{1})$}
                \State \textbf{break}
            \EndIf
            \State $\mathcal{R}(\hat{\mathbb{S}}, \mathbb{P}_{1}) \gets \mathcal{R}(\hat{\mathbb{S}}, \mathbb{P}_{1}) \cup Pre$
        \EndWhile
    \end{algorithmic}
\end{algorithm}

To compute the operations for Table~\ref{table:constructing_tlts}, we need to compute the BRS for system~\eqref{chapter_preliminaries:eq:dynamics_model} using $\mathcal{HZ}s$. Since we are dealing with discrete-time linear dynamics, we can use the following formula the formula developed in \cite{Zhang2023Backward} for computing the predecessor set~\eqref{eq:predecessor_set}:

The predecessor set from a target set $\mathcal{T} = \{ \vect{c}_{t}, G_{t}^{c}, G_{t}^{b}, A_{t}^{c}, A_{t}^{b}, \vect{b}_{t} \} \subseteq \mathbb{R}^{n_{x}}$ from the augmented state space $\hat{\mathbb{S}} = \{ \vect{c}_{s}, G_{s}^{c}, G_{s}^{b}, A_{s}^{c}, A_{s}^{b}, \vect{b}_{s} \} \subseteq \mathbb{R}^{n_{x} + n_{u}}$ is defined as $\mathcal{B} = \{ \vect{c}_{b}, G_{b}^{c}, G_{b}^{b}, A_{b}^{c}, A_{b}^{b}, \vect{b}_{b} \}$, where
\begin{equation}
    \begin{split}
        & G_{b}^{c} = 
        \begin{bmatrix}
            G_{s}^{c}[1:n_{x}, :] & \vect{0}
        \end{bmatrix}, \hspace{0.4cm}
        A_{b}^{c} = 
        \begin{bmatrix}
            A_{s}^{c} & \vect{0} \\
            \vect{0}  & A_{t}^{c} \\
            [A~B] G_{s}^{c} & -G_{t}^{c}
        \end{bmatrix}, \\        
        & G_{b}^{b} = 
        \begin{bmatrix}
            G_{s}^{b}[1:n_{x}, :] & \vect{0}
        \end{bmatrix}, \hspace{0.4cm}
        A_{b}^{b} = 
        \begin{bmatrix}
            A_{s}^{b} & \vect{0} \\
            \vect{0}  & A_{t}^{b} \\
            [A~B] G_{s}^{b} & -G_{t}^{b}
        \end{bmatrix}, \\
        & \vect{c}_{b} = \vect{c}_{s}[1:n_{x}, :], \hspace{1.6cm}        
        \vect{b}_{b} =
        \begin{bmatrix}
            \vect{b}_{s} \\
            \vect{b}_{t} \\
            \vect{c}_{t} - [A~B]\vect{c}_{s}
        \end{bmatrix}. \\
    \end{split}
    \label{eq:hz_brs}
\end{equation}

While a linear model oversimplifies a system's dynamics, it can be avoided in practice. To do so, we define the state space as an $\mathcal{HZ}$, which allows us to set boundaries, as the $\mathcal{HZ}$ is the collection of multiple constrained zonotopes that are, by definition, bounded. Therefore, if we consider for example a vehicle, its velocity can be restricted as needed in different regions. This will allow us to emulate the behaviour of a piecewise-linear model.

A notable characteristic of this approach is how it restricts the BRS within the state space. Typically, the computation of reachable sets utilizing the commonly known formulas in \cite[Chapter~11]{Borelli2017Predictive} for the successor and predecessor sets includes computing all possible combinations of next or previous states and then intersect those with the state space to ensure that the reachable set is always contained within the state space. Formula \eqref{eq:hz_brs}, directly integrates this behavior within it, and there is no need for any additional computations to achieve that.

\begin{example}
    For a more intuitive understanding consider the portion $\varphi_{N} = (true \, \underline{\textit{U}} \, p_{1})$ of the LTL~\eqref{eq:ltl_example_pnf}. A high-level description of using this approach to construct this specific TLT is laid out in Algorithm~\ref{alg:construct_tlt}.

    The algorithm takes as input the LTL, the augmented state space, the target space, as well as the linear system's dynamics and outputs the root set node of the TLT, which in this case is the infinite horizon BRS. The hybrid zonotopes $\hat{\mathbb{S}}$ and $\mathbb{P}_{1}$ in this case correspond to the set of states satisfying the \textit{true} and $p_{1}$ atomic propositions respectively. The core of the algorithm involves the computation of this infinite horizon BRS starting from the target set. The process involves computing and combining predecessor sets of predecessor sets originating from the target $\mathbb{P}_{1}$ until it finally converges (no new states are added), which means the infinite horizon BRS has been computed.
    Finally, as shown in~\cite{Gao2021Temporal}, the successful construction of the TLT tells which states can satisfy the specification. 
\end{example}

    \section{SIMULATED PARKING SCENARIO}
\label{sec:simulated_parking_scenario}
A parking mission typically entails the navigation of vehicles within non-trivial environments, often resulting in non-convex and disjoint sets. Since our work focuses on demonstrating the method's capability to effectively address such spatial complexities, the parking scenario is a fitting case study to demonstrate this work.

\subsection{Vehicle Model}

We define the following a discrete-time linear dynamics model to describe the behavior of a vehicle
\begin{equation}
    \begin{split}
        \vect{x}_{\scriptscriptstyle t+1} =
        \underbrace{
        \begin{bmatrix}
            1 & 0 & dt &  0 \\
            0 & 1 &  0 & dt \\
            0 & 0 &  1 &  0 \\
            0 & 0 &  0 &  1
        \end{bmatrix}}_{A}
        \underbrace{
        \begin{bmatrix}
            x \\
            y \\
            v_{x} \\
            v_{y}
        \end{bmatrix}}_{\vect{x}_{\scriptscriptstyle t}}
        +
        \underbrace{
        \begin{bmatrix}
            0  & 0 \\
            0  & 0 \\
            dt & 0 \\
            0  & dt
        \end{bmatrix}}_{B}
        \underbrace{
        \begin{bmatrix}
            a_{x} \\
            a_{y}
        \end{bmatrix}}_{\vect{u}_{\scriptscriptstyle t}}.
    \end{split}
    \label{eq:linear_model}
\end{equation}

The state vector describes the vehicle's position and velocity in the $x-y$ plane, while the input vector is its acceleration. For this work we use a sampling time of $dt = 0.1 s$.

\begin{figure*}[ht]
    \centering
    \begin{subfigure}{0.3\textwidth}
        \includegraphics[width=1.0\textwidth]{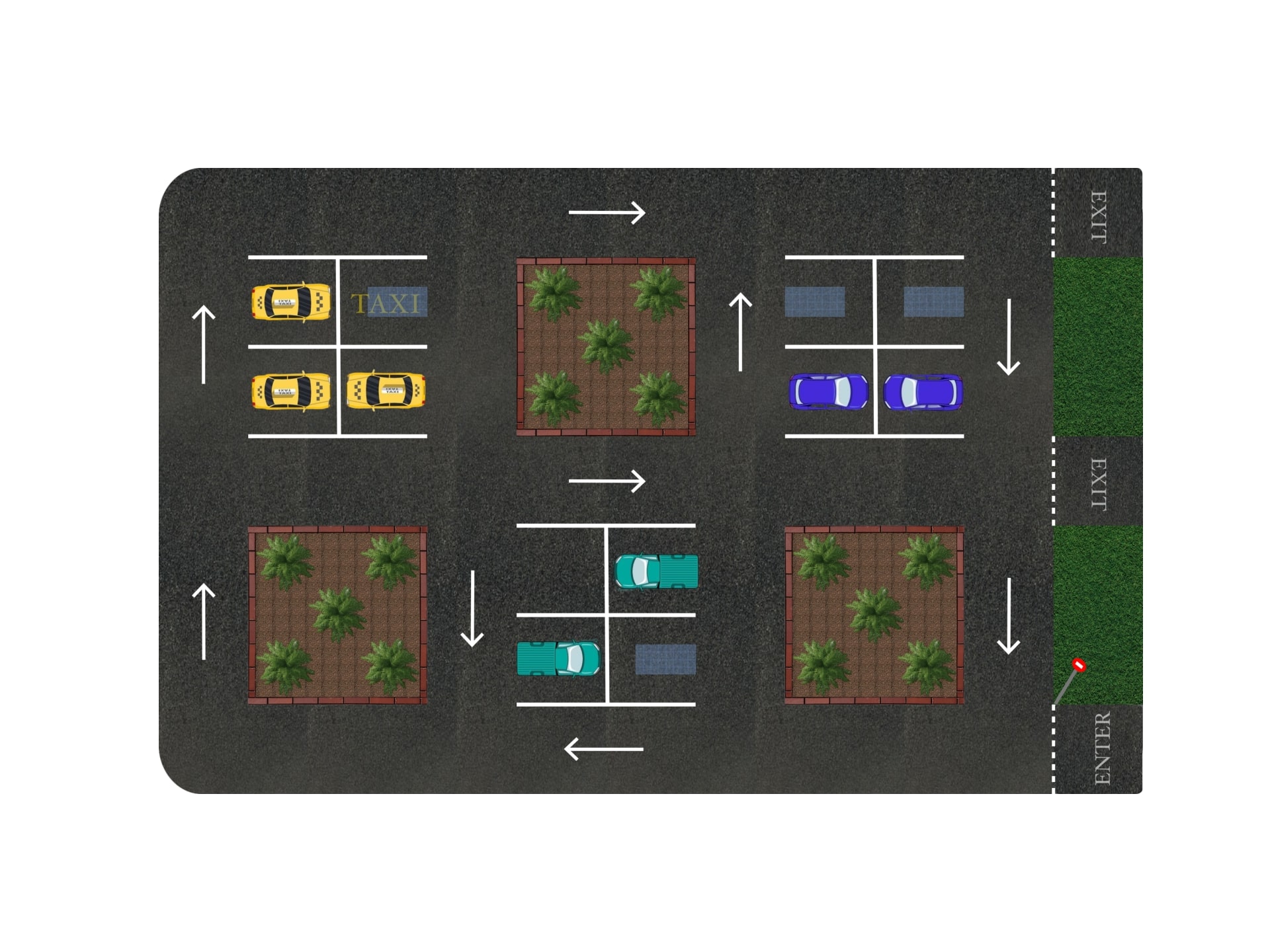}
        \vspace{-1.0cm}
        \captionsetup{margin={0.3cm, 0cm}}
        \caption{$\mathcal{R}(\mathbb{S}, \mathbb{P}_{1-5}, t = 0)$}
        \label{fig:results_0}
      \end{subfigure}
    \hfill
    \begin{subfigure}{0.3\textwidth}
        \includegraphics[width=1.0\textwidth]{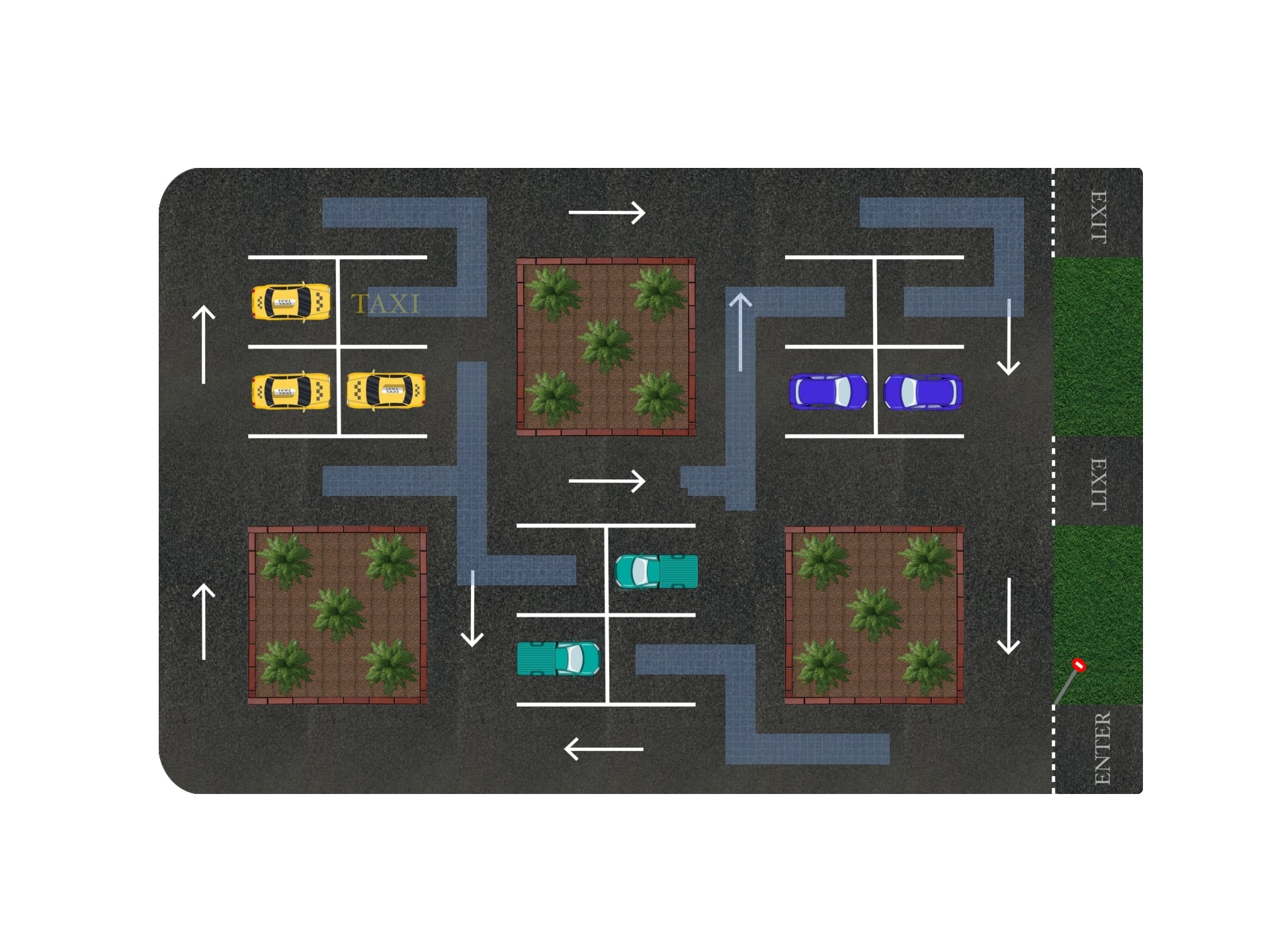}
        \vspace{-1.0cm}
        \captionsetup{margin={0.3cm, 0cm}}
        \caption{$\mathcal{R}(\mathbb{S}, \mathbb{P}_{1-5}, t = 30)$}
        \label{fig:results_30}
      \end{subfigure}
    \hfill
    \begin{subfigure}{0.3\textwidth}
        \includegraphics[width=1.0\textwidth]{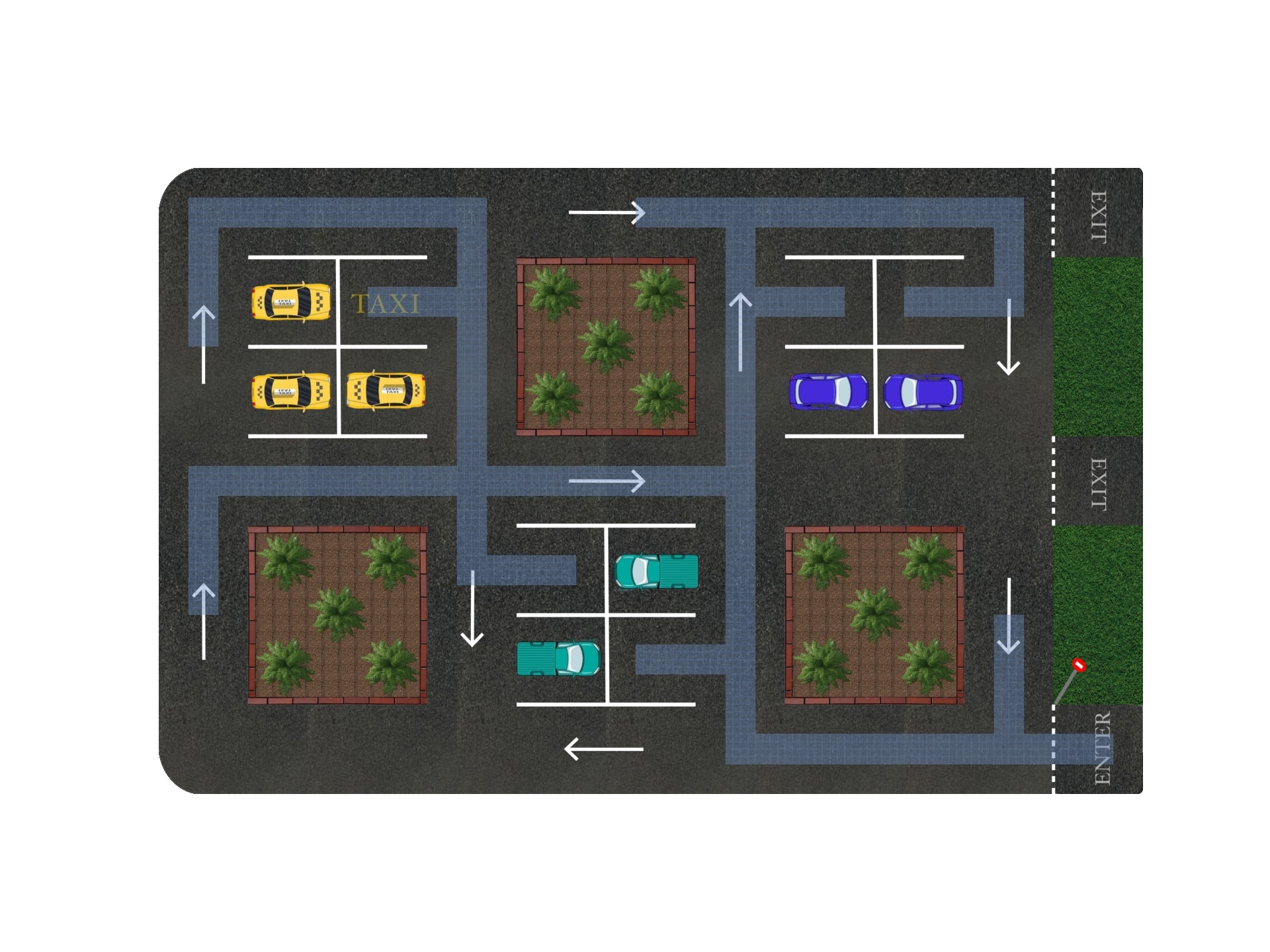}
        \vspace{-1.0cm}
        \captionsetup{margin={0.3cm, 0cm}}
        \caption{$\mathcal{R}(\mathbb{S}, \mathbb{P}_{1-5}, t = 60)$}
        \label{fig:results_60}
    \end{subfigure}
    \medskip
    \begin{subfigure}{0.3\textwidth}
        \includegraphics[width=1.0\textwidth]{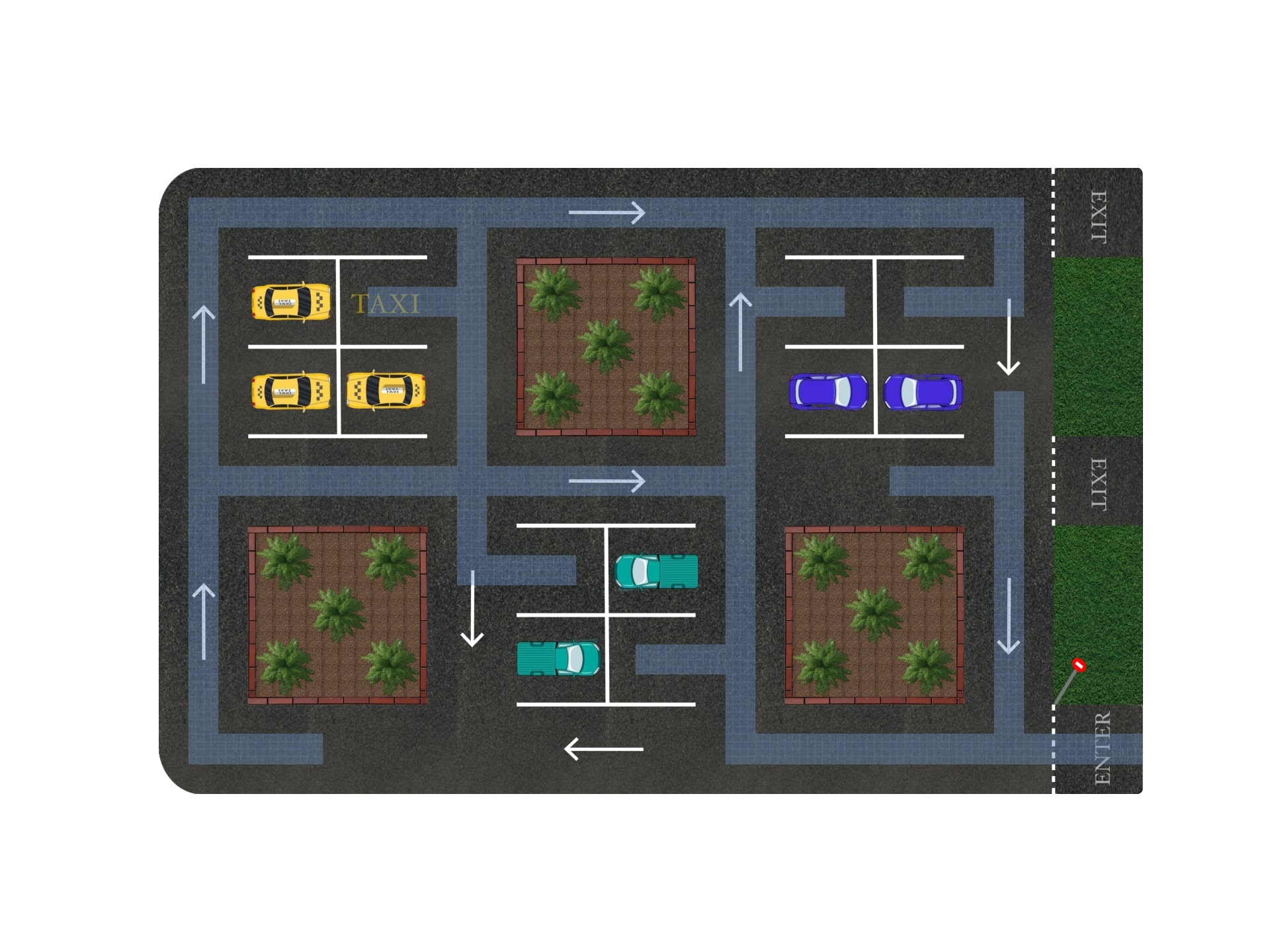}
        \vspace{-1.0cm}
        \captionsetup{margin={0.3cm, 0cm}}
        \caption{$\mathcal{R}(\mathbb{S}, \mathbb{P}_{1-5}, t = 90)$}
        \label{fig:results_90}
      \end{subfigure}
    \hfill
    \begin{subfigure}{0.3\textwidth}
        \includegraphics[width=1.0\textwidth]{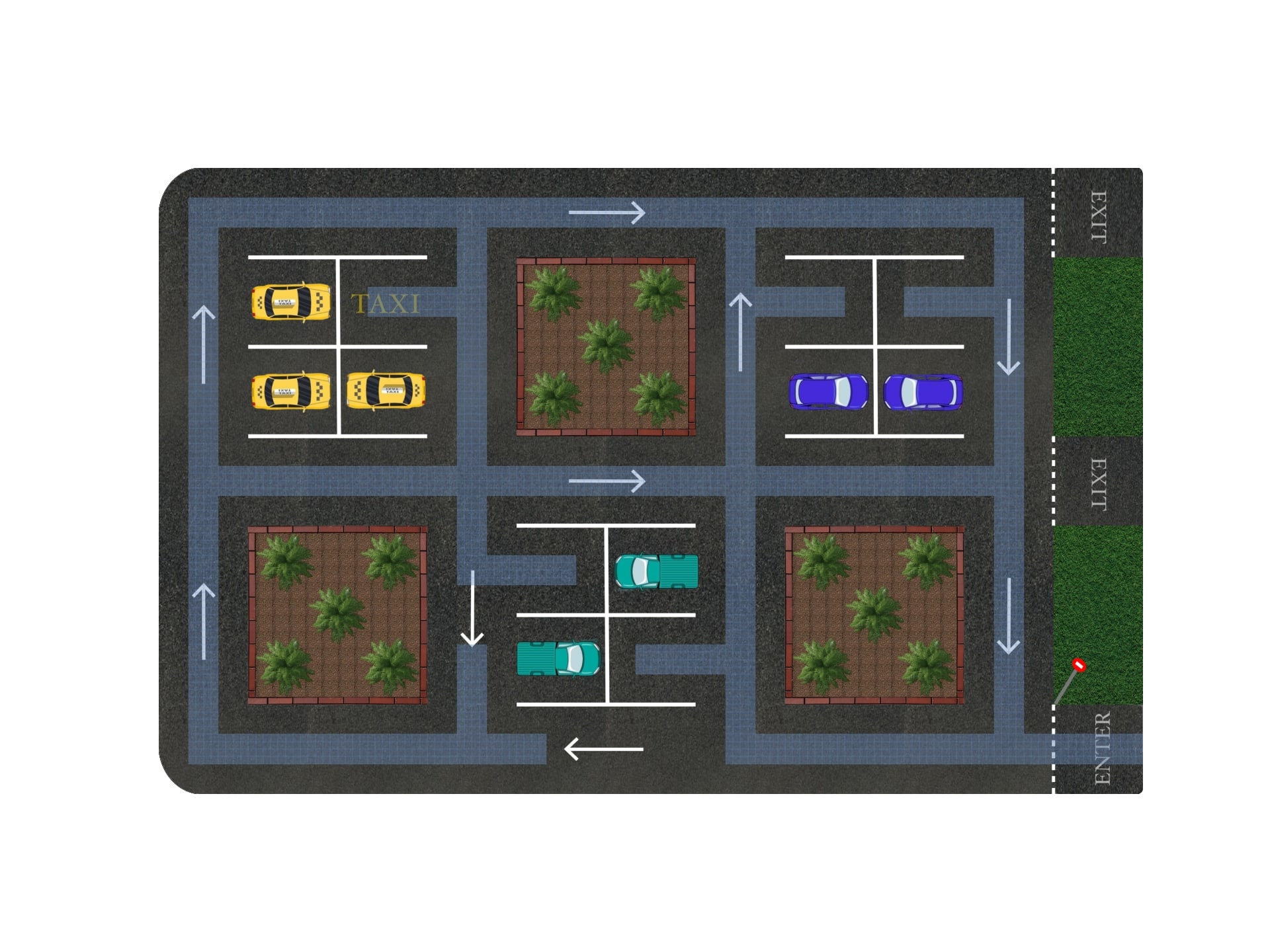}
        \vspace{-1.0cm}
        \captionsetup{margin={0.3cm, 0cm}}
        \caption{$\mathcal{R}(\mathbb{S}, \mathbb{P}_{1-5}, t = 120)$}
        \label{fig:results_120}
      \end{subfigure}
    \hfill
    \begin{subfigure}{0.3\textwidth}
        \includegraphics[width=1.0\textwidth]{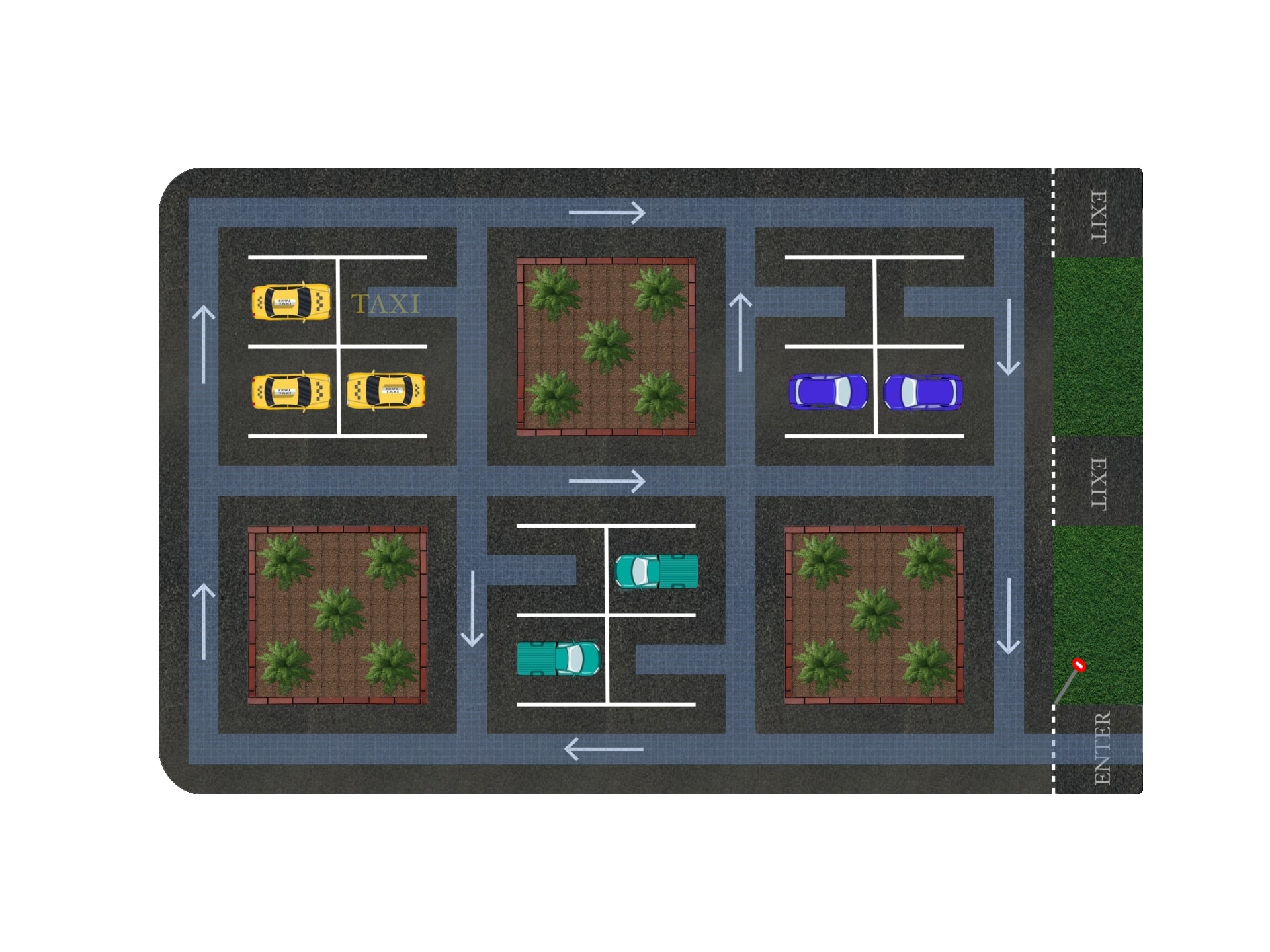}
        \vspace{-1.0cm}
        \captionsetup{margin={0.3cm, 0cm}}
        \caption{$\mathcal{R}(\mathbb{S}, \mathbb{P}_{1-5}, t = 150)$}
        \label{fig:results_150}
    \end{subfigure}
    \caption{Evolution of BRS from parking spots}
    \label{fig:results}
\end{figure*}

\subsection{Parking Environment}

All simulations in this work are conducted within the environment in Figure~\ref{fig:results}. The linear model in \eqref{eq:linear_model} does not capture the the vehicle's orientation and velocity constraints. To compensate for that, one can encode such information within the state and input spaces themselves.

Consider, for example, the top and bottom horizontal lanes in a sub-figure of Figure~\ref{fig:results}; the velocity and acceleration constraints, as well as the traffic direction rules, can be described by augmenting the state space with the input space as $\hat{\vect{x}}_{\scriptscriptstyle k} = \begin{bmatrix} \vect{x}_{\scriptscriptstyle k}^{T} & \vect{u}_{\scriptscriptstyle k}^{T} \end{bmatrix}^{T}$ and then using the following $\mathcal{HZ} = \{\vect{0}, G^{c}, G^{b}, \vect{0}, \vect{0}, \vect{0}\}$, where $G^{c} = \textit{diag(1.4, 0.05, 0.5, 1, 0.1, 0.1)}, ~ G^{b} = [0~0.9~0.5~0~0~0]^{T}$.

The continuous generators of the $\mathcal{HZ}$ form a hyperrectangle whose size in each dimension is double the corresponding diagonal element. In other words, a road lane with a size of ($2.8 \times 0.1) m^2$ is formed. In that road section the velocities $v_{x}, v_{y}$ are constrained in $([-0.5, 0.5], ~[-1.0, 1.0])~ m/s$ respectively, and the acceleration in $[-0.1, 0.1]~m/s^{2}$. To duplicate this road as well as adjust its position and velocity parameters we use its binary generators. Since there is one binary generator, the $\mathcal{HZ}$ is equivalent to the union of these two constrained zonotopes $\mathcal{CZ}_{1} = \{G^{b}, G^{c}, \vect{0}, \vect{0} \}, ~\mathcal{CZ}_{2} = \{-G^{b}, G_{2}, \vect{0}, \vect{0} \}$.

Each constrained zonotope represents one of the two horizontal lanes and forces the vehicle to abide by the traffic direction rules. Following similar reasoning, the rest of the augmented state space can be defined and collectively described using a single $\mathcal{HZ}$ and is denoted as $\hat{\mathbb{S}}$.

\subsection{Constructing the TLTs}

The case study considered in this work is to provide guarantees for an autonomous vehicle parking task. This mission was introduced in Example~\ref{example:tlt}. Following that example, the resulting LTL specification is restated here.
\begin{equation}
    \begin{split}
        \varphi = (\lozenge p_{1}) \lor (\lozenge p_{2}) \lor (\lozenge p_{3}) \lor (\lozenge p_{4}) \lor (\lozenge p_{5}), \\
        p_{i} = \{ \vect{x} \in \mathbb{S} | \vect{x} \in \mathbb{P}_{i} \}, \hspace{0.3cm} 
        i = \{1, 2, ..., 5\}. \hspace{0.35cm}
    \end{split}
    \label{eq:case_study_ltl}
\end{equation}
 
An equivalent TLT capturing the formula (\ref{eq:case_study_ltl}) is shown in Figure~\ref{fig:example_tlt}. The construction of this TLT requires the computation of BRSs, as well as their union. Starting with a single branch for parking spot $\mathbb{P}_{i}$, $\mathcal{R}(\mathbb{S}, \mathbb{P}_{i})$ needs to be computed in order to find all $(x, y, v_{x}, v_{y})$ states for which the vehicle can reach that parking spot. Using equation~\eqref{eq:hz_brs}, we need to augment the state space to include the car's acceleration and thus use the augmented state space $\hat{\mathbb{S}}$ instead.

Given that the environment is static, the state space can be directly defined such that it only contains the roads and excludes any obstacles. In addition, to account for the vehicle's dimensionality, the state space is conservatively shrunk towards the center of each lane, as the reachable set computations assume a point mass object.

    \section{RESULTS}
\label{sec:results}

To assess the effectiveness of the combination of $\mathcal{HZ}$ reachability analysis and temporal logic, we now proceed to compute one by one all the set nodes of Figure~\ref{fig:example_tlt}. 

Starting with the set nodes $\mathbb{P}_{1-5} \coloneqq \{\mathbb{P}_{1}, \mathbb{P}_{2}, ..., \mathbb{P}_{5}\}$, these represent the targets of our LTL; that is the five available parking spots. The set nodes $\mathcal{R}(\mathbb{S}, \mathbb{P}_{1}), \mathcal{R}(\mathbb{S}, \mathbb{P}_{2}), ..., \mathcal{R}(\mathbb{S}, \mathbb{P}_{5})$ correspond to the BRS for each available parking spot and can be computed independently from each other in parallel to immediately generate the root set node of the tree.

The evolution of the computation of the root node is shown in Figure~\ref{fig:results}. There, we see its evolution in increments of $30$ steps starting from the original target sets in Figure~\ref{fig:results_0} until the set finally converges after $150$ steps to the one shown in Figure~\ref{fig:results_150}. For visualization purposes, the blue curve designates all positions $(x, y)$ for which there is at least one velocity state that can satisfy the goal.

Right from the first time instance, we see how the $\mathcal{HZ}$ is capable of handling the five disjoint target sets simultaneously. The ability of the $\mathcal{HZ}$ to handle complex geometric representations is further verified during the next instances, where, for example, Figure~\ref{fig:results_30} shows how the BRS, starting from the fourth parking spot, splits into two distinct paths. Then, we see how they all merge and grow together to cover the rest of the space. As expected, the final BRS covers the entire state space, aside the two exits of the parking lot, as a car is not allowed to move opposite to the traffic direction.

Earlier, we claimed that the simplicity of the dynamics model is not necessarily of great significance in describing complex tasks because the computation of BRSs using $\mathcal{HZ}$s allows us to capture the input space information by augmenting the state space and thus enforcing the speed limit and traffic direction rules. This claim is indeed supported by the results since the evolution of the BRS always goes against the traffic direction, which is expected since we are computing a BRS and propagating backwards in time.

Sample computational times for the construction of the TLT both for the 4D and also for a simplified 2D version are in the range of $8.56 s$ and $3.94 s$, respectively. These results serve as a preliminary indication of the efficiency potential of $\mathcal{HZ}$s.

    \section{CONCLUSION}
\label{sec:conclusion}

Ensuring the safety of autonomous vehicles throughout their entire operation poses a challenging yet vital task that requires our attention. In this work, we have shown how LTL can be used to formally specify parking and navigation missions in dynamic environments and how one might use $\mathcal{HZ}$ reachability analysis to guarantee safety during these missions.

Although most works employing zonotope reachability analysis use them for their notable computational efficiency, few works address non-trivial spaces. This higher level of complexity in the problem is enabled through the use of temporal logic formalism, and the use of $\mathcal{HZ}$s allows us to accelerate the computation of the reachable sets. 

Extensions of this work include applying this method to more abstract goals, such as high-level planning in smart cities, and exploiting the hybrid nature of this work through distributed system problems. Other interesting avenues are the extension of this approach to deal with dynamic environments and temporal tasks utilizing the potential of $\mathcal{HZ}$s for an online solution. Finally, it is important to perform formal control synthesis using the computed TLT in this work.
    \balance

    \addtolength{\textheight}{-12cm}

    \bibliographystyle{IEEEtran}
    \bibliography{IEEEabrv,bibliography.bib}
    
\end{document}